# Analysis of Fission Matrix Databases using Temperature Profiles obtained from High-Fidelity Multiphysics Simulations

Maximiliano Dalinger, Elia Merzari, Saya Lee, and Alex Nellis

*Pennsylvania State University, Hallowell Building, State College, PA, mgd5394@psu.edu, ebm5351@psu.edu, sayalee@psu.edu, vxn5069@psu.edu*

## INTRODUCTION

The Monte Carlo method uses continuous-energy cross sections, angle variables, and exact geometry modeling, making it the gold standard for neutronic calculations. However, to achieve results with lower uncertainty, a large number of particles are needed, resulting in a high computational cost. This makes Monte Carlo less attractive for situations requiring temperature feedback or transient calculations [1]. The Fission Matrix (FM) method is a possible technique for quickly performing neutronics calculations. It relies on pre-calculated databases generated from Monte Carlo simulations to obtain an accurate neutronic solution at relatively low computational cost [1,2].

A temperature interpolation method can be used to generate an FM with the desired temperature profile. Previous works used Fission Matrix Databases (FMDBs) generated with uniform and non-uniform fuel temperature profiles to study their effects on the multiplication factor and fission source distribution [3] for the Molten Salt Fast Reactor (MSFR) [4]. However, the temperature profiles obtained in thermal hydraulics simulations of the MSFR showed a different shape [8] from those analyzed in [3]. In this paper, we use Cardinal to perform high-fidelity multiphysics simulations coupling neutronics and thermal hydraulics to obtain more accurate temperature profiles. These are used to generate a new set of FMDBs. The influence of the temperature profile used to generate the FMDB on the multiplication factor and fission source distribution is studied.

## METHODOLOGY

### Cardinal

Cardinal [5] is an open-source application that wraps OpenMC and NekRS codes within the MOOSE framework. It uses the MOOSE data transfer implementation to perform high-fidelity coupled neutronic and thermal hydraulics calculations. It can also be coupled with any MOOSE-based application, enabling broad multiphysics capabilities.

### The Fission Matrix Method

A fission matrix is an *n x n* matrix $A$ which corresponds to a discretization of the reactor geometry into $n$ spatial cells. Each element of this matrix, $a_{ij}$, represents the number of neutrons generated in cell $i$ from a fission neutron born in cell $j$. The fundamental eigenvector of $A$ is the fission source distribution $\vec{F}$, and the fundamental eigenvalue is the multiplication factor $k_{eff}$. The FM Method formulation [6] is shown in Equations 1.1 and 1.2. Solving the FM means obtaining the $k_{eff}$ and the fission source distribution.

$$\vec{F} = \frac{1}{k_{eff}} A \vec{F} \qquad (1.1)$$

$$F_i = \frac{1}{k_{eff}} \sum_{j=1}^{N} a_{ij} F_j \qquad (1.2)$$

## COMPUTATIONAL MODELS

The high-fidelity multiphysics model used to generate the temperature profiles is based on reference [7]. Here, it is briefly explained for completeness.

### OpenMC MSFR Model

The OpenMC model was generated using a simplified cylindrical geometry using CSG techniques. The liquid fuel is LiF-ThF$_4$-$^{233}$UF$_4$, and the breeder blanket contains LiF-$^{232}$ThF$_4$. The blanket is surrounded by a boron carbide (B$_4$C) layer to absorb any remaining neutrons. Radial and axial reflectors surround the entire core and are constructed of HT9 stainless steel. Figure 1 shows the OpenMC geometry colored by material (A and B) and cell (C and D), where red represents the liquid fuel, blue is the fertile blanket, green is the absorber, and grey is the stainless-steel reflector. The fuel region is discretized to account for the change in fuel temperature and density. Temperature and density feedback are applied exclusively to the fuel region, while other components are always at 973 K. The entire geometry is surrounded by a void cell with vacuum boundary conditions. Simulations used 1000000 particles, with 20 inactive batches and 100 total batches, obtained from a Shannon Entropy analysis, which resulted in uncertainties below 12.5 pcm for the effective multiplication factor $k_{eff}$. The cross-section library used was ENDF/B-VIII.0. The OpenMC model has been verified against Serpent in [7], showing a good agreement for the neutron flux distribution.

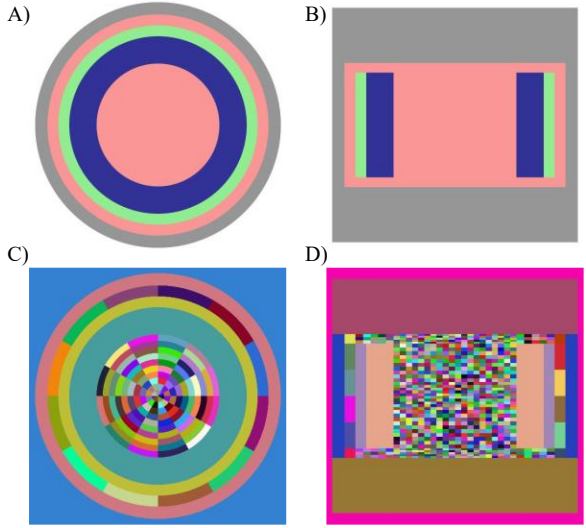

Fig. 1. OpenMC model of the MSFR. A) upper view colored by material, B) lateral view colored by material, C) upper view colored by cell, and D) lateral view colored by cell.

**NekRS MSFR Model**

NekRS solves fluid mass, momentum, and energy conservation using a RANS $k-\tau$ turbulence model. It receives the volumetric heat source calculated in OpenMC. Figure 2 shows the three-dimensional geometry of the MSFR primary loop used in the calculations, generated with Gmsh. The model uses simplified geometry with a continuous inlet/outlet and does not model the pumps or heat exchangers. The separation between the inlet and the outlet is 50 cm. The boundary conditions for the energy equation are a uniform inlet temperature of 898 K and adiabatic walls; the code calculates the outlet temperature. Regarding the momentum equation, the boundary conditions are a parabolic velocity profile with a mean non-dimensional value of 1 at the inlet, non-slip at the remaining walls, and the code calculates the outlet velocity. For turbulence variables $k$ and $\tau$, parabolic profiles were used at the inlet with mean non-dimensional values of 0.01 and 0.1, respectively. Table I lists the relevant non-dimensional parameters used in the simulations. The mesh has $1.35*10^6$ hexahedral elements. A polynomial order 3 was selected, resulting in about $8.68*10^7$ Gauss-Lobatto-Legendre (GLL) quadrature points. This resolution was found adequate for the problem. A 2D RANS version of the NekRS model was verified against OpenFOAM in [7], showing a good agreement for the velocity field.

Table I. Nondimensional parameters used for NekRS coupled calculations.

| Nondimensional Parameter | Symbol | Value |
|---|---|---|
| Reynolds Number | $Re$ | 30672 |
| Prandtl Number | $Pr$ | 16 |
| Peclet Number | $Pe$ | 491286 |
| Turbulent Prandtl Number | $Pr_t$ | 1 |

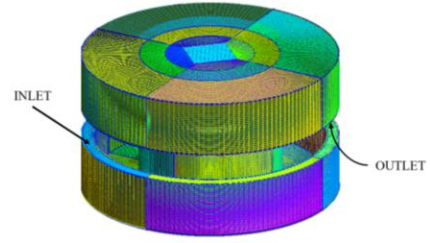

Fig. 2. NekRS mesh of the MSFR.

**Cardinal Coupling Methodology**

Figure 3 presents the coupling procedure used in Cardinal, often called the Picard iteration "in time" [5]. OpenMC runs a k-eigenvalue calculation using an initial uniform temperature and density. It calculates the heat source $q'''_{fis}$. Cardinal receives this information and sends it to NekRS. Then NekRS performs $N$ calculations of time step $\Delta t_{nek}$ to calculate fluid temperature and velocity distributions. NekRS sends the fluid temperature. Cardinal calculates the fuel density [8] and sends both fuel temperature and density. Temperatures and numerical densities are updated in OpenMC, and the process is repeated until convergence. Convergence is reached when the maximum, average, and outlet temperatures reach steady-state values. For the present work, $N = 4000$ and $\Delta t_{nek} = 2.0*10^{-4}$. Therefore, the OpenMC model receives feedback for the fuel temperature and density. This model is used to generate the temperature profiles for the FM database generation stage.

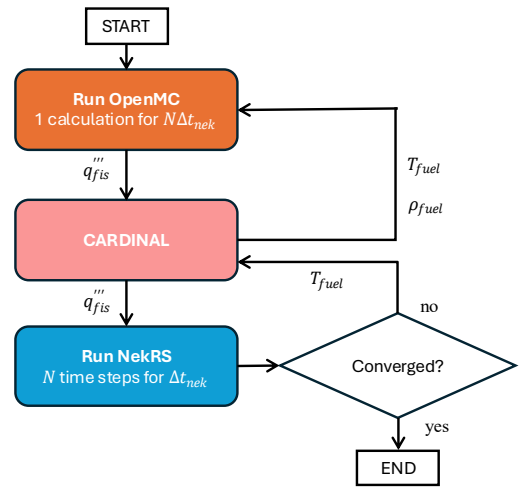

Fig. 3. Cardinal coupling procedure.

**Fission Matrix Databases Generation**

FMs are obtained from Monte Carlo simulations for a specific reactor state, i.e., a particular fuel temperature distribution. To accurately represent the reactor state, which differs from that used in Monte Carlo simulations, multiple FMs are required, known as Fission Matrix Databases (FMDBs). The FM representative of the reactor's actual state is derived from the FMDBs. In the present work, FMDBs are generated from a set of fuel temperature profiles calculated in Cardinal. The Monte Carlo code Serpent 2.2 was used to generate the FMDBs. The 3D model of the MSFR resembles the OpenMC model. In all cases, Serpent calculations used 50000 particles, with 1000 active cycles and 20 inactive cycles, resulting in uncertainties below 7.1 pcm for $k_{eff}$ and a maximum RMS of 0.9% for the fission source distribution. The cross-section library used was ENDF/B-VII.1. FMDBs were calculated with a 13x13x13 Cartesian discretization.

**FM Linear Interpolation**

Matrix $A$ is calculated for a particular fuel temperature distribution. To do this, each element $a_{ij}$ is linearly interpolated between the two nearest FMDBs corresponding to the nearest cell temperatures. Interpolations follow Equation 2, where $d$ and $d+1$ are the nearest database indexes, and $T_i$ is the destination cell temperature. This procedure has been previously analyzed [2].

$$a_{ij} = \tilde{a}_{ij}^{(d+1)}\left(\frac{T_i - T_i^{(d)}}{T_i^{(d+1)} - T_i^{(d)}}\right) + \tilde{a}_{ij}^{(d)}\left(1 - \frac{T_i - T_i^{(d)}}{T_i^{(d+1)} - T_i^{(d)}}\right) \quad (2)$$

**Fission Matrix Model**

FM calculations are performed using a C++ script. It loads the FMDBs, receives the desired temperature profile, generates the current FM by linear interpolation, and solves it, obtaining the corresponding $k_{eff}$ and fission source distribution. The model uses the same discretization as the databases. The script also enables the loading of reference results, which is useful to compare the $k_{eff}$ and fission source. Uncertainty is propagated following reference [9]. It is planned to add the script to MOOSE in the future.

**RESULTS**

**Temperature Profiles Generation**

The supercomputer Frontier was used for Cardinal calculations. Calculations were performed for approximately 100 convective units, and then time-averaging was performed for approximately 400 more convective units. Then, temperature profiles were azimuthally averaged and transferred into a MOOSE Cartesian mesh of 13x13x13 to generate the desired profiles.

Simulations to generate database temperature profiles used a modified inlet temperature boundary condition, keeping all other parameters constant. Inlet temperatures ranged from 700 K to 1300 K in 100 K steps, yielding seven profiles. Figure 4 shows the temperature profile for an inlet temperature of 800 K obtained in Cardinal, along with the values after transfer to the Cartesian mesh. Cells with a white color do not receive a temperature value from Cardinal because they do not have fuel.

Two more temperature profiles were generated in Cardinal. The first used an inlet temperature of 898 K, as the reference MSFR Cardinal's model [7]. The second changed the mass flow rate to 1700 kg/s. These profiles were used as test profiles to solve the FM problem.

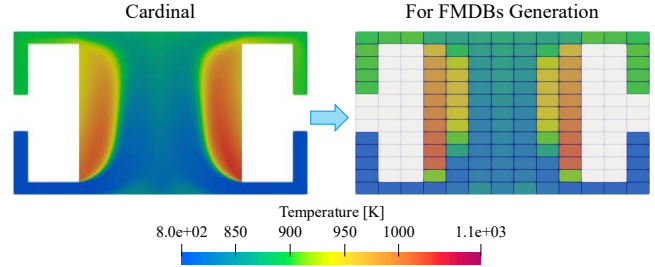

Fig. 4. Temperature profile obtained in Cardinal (left) and in the Cartesian mesh (right), for an inlet temperature of 800 K.

**Fission Matrix Results**

Two sets of FMDBs were analyzed: 1) the set generated with Cardinal temperature profiles, and 2) a set generated using uniform temperature profiles, for temperatures ranging from 800 K to 1200 K in steps of 100 K. Simulations were performed in the FM model and compared with Serpent results obtained using the same test temperature profile. Differences between the FM model and Serpent are presented for the $k_{eff}$, the fission source root-mean-square (RMS) and maximum difference. Tables II and III present the differences using test profiles 1 and 2, respectively. As can be seen, the differences are lower when using the FMDBs obtained using Cardinal temperatures.

Table II: Results using test profile 1.

| FMDB | $k_{eff}$ Difference [pcm] | Fission Source RMS Difference [%] | Fission Source Maximum Difference [%] |
|---|---|---|---|
| 1-Cardinal | -3.1 ± 33.4 | 1.8 ± 1.7 | 18.5 ± 10.3 |
| 2-Uniform | -23.9 ± 33.5 | 1.9 ± 1.7 | 22.8 ± 8.6 |

Table III: Results using test profile 2.

| FMDB | $k_{eff}$ Difference [pcm] | Fission Source RMS Difference [%] | Fission Source Maximum Difference [%] |
|---|---|---|---|
| 1-Cardinal | 7.4 ± 32.9 | 1.8 ± 1.7 | 17.4 ± 9.7 |
| 2-Uniform | -14.7 ± 33.1 | 1.8 ± 1.7 | 24.6 ± 8.6 |

Figure 5 presents a slice of the fission source distribution obtained with the FM model using test profile 1 and Cardinal's FMDB, at index 7 corresponding to the center of the core. Figure 6 presents the difference between the above-mentioned profile and the fission source from Serpent. Figure 7 presents the difference in the fission source between Serpent and the FM model using the uniform's FMDB for the same test temperature profile. As shown in Figure 7, the differences are negative at the bottom of the core and positive at the top. This is the influence of the temperature profiles used to generate the FMDBs. Therefore, the best approach is to use FMDBs generated with temperature profiles similar to those expected when solving the FM model.

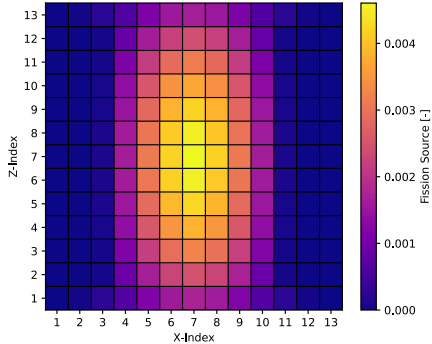

Fig. 5. XZ slice of the fission source distribution at index 7, using Cardinal's FMDB and test profile 1.

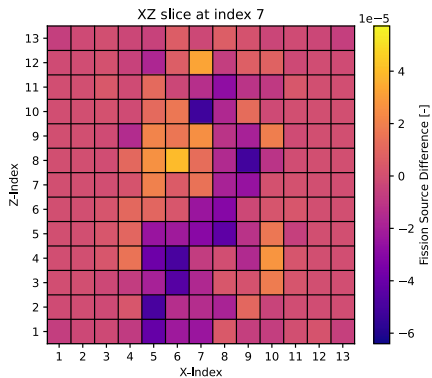

Fig. 6. XZ slice of the fission source difference from Serpent at index 7, using Cardinal's FMDB and test profile 1.

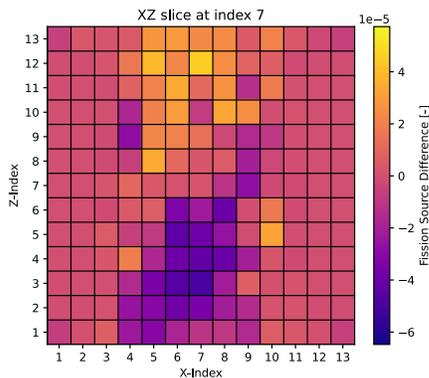

Fig. 7. XZ slice of the fission source difference from Serpent at index 7, using uniform's FMDB and test profile 1.

## CONCLUSIONS

In the MSFR, the temperature profiles are complex due to the internal recirculation and stagnation regions inside the reactor. Multiphysics simulations coupling neutronics and thermal hydraulics were performed in Cardinal to obtain temperature results, which were used to generate a set of FMDBs. The FM model was used to perform simulations using different test temperature profiles. Two FMDBs were used: the first obtained from Cardinal temperature profiles, and the other from uniform temperatures. Results showed that there is an improvement in the $k_{eff}$ and fission source distribution when the temperatures used to generate the FMDB are like the profiles used in the FM model.

Future analysis proposes analyzing other temperature profiles for testing and generating the FMDB, as well as the discretization used in the FM model.

## ACKNOWLEDGMENTS

This research was in part performed using funding received from the U.S. NRC, Grant #31310022M0037.